LA-UR-11-12163

# Non-Demolition Dispersive Measurement of a Superconducting Qubit with a Microstrip SQUID Amplifier

G.P. Berman[a], D.I. Kamenev[a], D. Kinion[b], and V.I. Tsifrinovich[c]

[a]Theoretical Division and CNLS, Los Alamos National Laboratory, Los Alamos, NM 87545, USA

[b]Lawrence Livermore National Laboratory, Livermore, CA 94551, USA

[c]Department of Applied Physics, Polytechnic Institute of NYU, 6 MetroTech Center, Brooklyn, NY 11201, USA

## Abstract

We have studied the possibility of a single-shot non-demolition measurement of a superconducting qubit using a microstrip SQUID amplifier (MSA). The Johnson noise generated by all resistors in the MSA is taken into consideration. We show that a single-shot non-demolition measurement is possible with six photons in the measurement resonator. For a phase qubit inductively coupled to a measurement resonator we have obtained the expression for the mutual inductance required for measurement of the qubit state.

## 1. Introduction

A superconducting circuit containing Josephson junctions (JJ) has attracted enormous attention of researchers developing quantum computers [1-6]. This circuit can be considered as an artificial micrometer-size atom, which can be easily integrated into sophisticated electronic architectures. One of the main problems for any implementation of quantum computing is achieving a high fidelity measurement of the qubit computational states. To do this one needs a non-demolition quantum measurement [7]. A non-demolition measurement does not disturb the computational state. It allows one to integrate the signal and increase the signal-to-noise ratio (SNR) to that required for a high fidelity measurement.

Various schemes for non-demolition measurements of a superconducting qubit have been discussed in literature [8,9], but the only scheme implemented experimentally is the dispersive measurement designed in ref. [10]. In this method a qubit loop (QL) is coupled to a measurement resonator (MR) with a small number of photons. Due to the small number of photons and relatively large detuning between the MR and the QL, the computational quantum state of the QL remains undisturbed while the frequency of the MR becomes dependent on the qubit state. Experimental implementation of this scheme requires amplification of the MR signal. The inevitable noise generated by an amplifier became the main obstacle for a single-shot non-demolition measurement [7]. To the best of our knowledge, currently all experiments with a non-demolition dispersive measurement require multiple-shot averaging, which is unacceptable for quantum computation with many qubits.

To reduce the amplifier-generated noise one could use a low-noise microstrip SQUID amplifier (MSA) [11]. At frequencies below 1GHz MSAs have demonstrated a 25dB power gain with an almost quantum limited noise temperature of $2hf/k_B$ [12]. However, for a dispersive measurement one has to use an MSA in the GHz frequency region. This requires reduction of the size of the MSA input coil, which diminishes its coupling with the SQUID [11]. One could raise the input voltage of the MSA increasing the number of photons in MR but this causes a back reaction of the MR on the QL, and the dispersive measurement becomes a demolition measurement [13-15].

In the first part of our work we try to find a delicate balance between the reasonable MSA parameters and the number of photons in the MR. Applying the simple model of an MSA suggested in [16] and experimental data from the non-demolition dispersive measurement with a transmon qubit [17], we show that a single-shot non-demolition dispersive measurement is possible with six photons in the MR.

A phase qubit occupies a special place in the superconducting qubit zoo: it is especially adjusted for integration into the complicated quantum computer architecture [18]. To the best of our knowledge, the non-demolition dispersive measurement has not been implemented for the phase qubit. In the second part of this work, we consider a phase qubit inductively coupled to the MR and derive an expression for the QL-MR interaction constant. Based on this expression we formulate a condition on the mutual QL-MR inductance required for a single-shot non-demolition measurement of a phase qubit.

## 2. Dispersive measurement of a superconducting qubit with an MSA

We consider a measurement scheme similar to that used in [17], but containing an MSA. (See Fig. 1.) A driving electromagnetic field (measurement tone) of frequency, $f_m$, is applied to the MR, which is coupled to the QL. The frequency of the MR, $f_r$, equals the driving frequency, $f_m$, when the qubit is in its excited state, $|1\rangle$. If the qubit is in its ground state, $|0\rangle$, then $f_r \neq f_m$. The output MR voltage is amplified by the MSA, which is tuned to the driving frequency, $f_m$. It is clear that the output MR voltage, which is equal to the MSA input voltage, depends on the qubit state: for the qubit excited state it is expected to be much greater than for the qubit ground state.

After amplification, the output MSA voltage is mixed with the reference signal whose frequency is slightly different from the driving frequency, $f_m$, (heterodyne detection). The mixed signal whose frequency is much lower than the driving frequency is finally detected. Depending on the phase of the reference signal, the amplitude of the mixed signal can represent either in-phase or the quadrature signal.

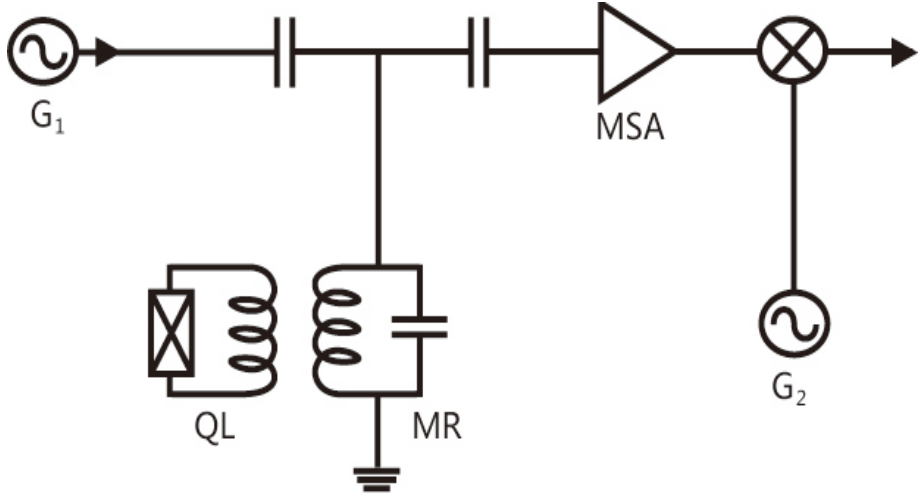

Fig. 1. Schematic of a dispersive measurement with an MSA and a heterodyne detection. $G_1$ is the generator of the driving electromagnetic field, $G_2$ is the generator of the reference signal (local generator), $\otimes$ indicates the mixer.

The non-demolition measurement of the qubit state is described by the Cavity-Bloch equations in the rotating frame [17]:

$$\begin{aligned}
d_t \langle a \rangle &= -i\Delta_{rm} \langle a \rangle - i\chi \langle a\sigma^z \rangle - i\varepsilon_m - (\kappa/2)\langle a \rangle, \\
d_t \langle \sigma^z \rangle &= -\gamma_1 \left(1 + \langle \sigma^z \rangle\right), \\
d_t \langle a\sigma^z \rangle &= -i\Delta_{rm} \langle a\sigma^z \rangle - i\chi \langle a \rangle - i\varepsilon_m \langle \sigma^z \rangle - \gamma_1 \langle a \rangle - (\gamma_1 + \kappa/2)\langle a\sigma^z \rangle.
\end{aligned} \quad (1)$$

Here $a$ is the annihilation operator which describes the electromagnetic field in the MR; $\Delta_{rm} = \omega_r - \omega_m$, is the difference between the unperturbed MR frequency and the driving frequency, $\omega_k = 2\pi f_k$ for any index "$k$"; $\chi$ is the MR frequency shift; $\sigma^z$ is the Pauli operator describing the qubit state; $\varepsilon_m$ is the amplitude of the driving field in frequency units; $\kappa$ is the MR decay constant, which is connected to the MR quality factor, $Q$, by the relation: $\kappa = \omega_r / Q$; and $\gamma_1$ is the decay constant for the QL, which is a reciprocal of the qubit relaxation time: $\gamma_1 = 1/T_1$. In these equations $\langle \sigma^z \rangle = 1$ for the excited qubit state. Below we put $\Delta_{rm} = -\chi$. The average number of photons in the resonator, $\langle n \rangle = \langle a^\dagger a \rangle$, is given by the equation:

$$d_t \langle n \rangle = -2\varepsilon_m \mathrm{Im}\langle a \rangle - \kappa \langle n \rangle. \quad (2)$$

The value, $\varepsilon_m = \kappa/2$, corresponds to one photon in the MR in the stationary resonance regime with no qubit relaxation $(\langle\sigma^z\rangle = 1, d_t\langle a\rangle = d_t\langle a\sigma^z\rangle = \gamma_1 = 0)$. We will consider a rectangular driving pulse: $\varepsilon_m = 0$ for $t < 0$, and $\varepsilon_m = \kappa/2$ for $0 < t < \tau$. With initial conditions:

$$\langle\sigma^z\rangle = \pm 1, \quad \langle a\rangle = 0, \quad \langle a\sigma^z\rangle = 0, \tag{3}$$

the Cavity-Bloch equations can be easily solved analytically. In particular, the stationary solutions with no qubit relaxation are given by the simple equations,

$$\langle a\rangle = -2i\varepsilon_m/\kappa \text{ for } \langle\sigma^z\rangle = 1,$$
$$\langle a\rangle = -2i\varepsilon_m/(\kappa - 4i\chi) \text{ for } \langle\sigma^z\rangle = -1. \tag{4}$$

Note that for $\langle\sigma^z\rangle = 1$ the stationary solution is imaginary, i.e. at the output of the MR we have only the quadrature voltage, $V$, which is proportional to $\text{Im}\langle a\rangle$:

$$V = (hf_r\kappa Z)^{1/2} \text{Im}\langle a\rangle. \tag{5}$$

Here $Z$ is the characteristic impedance of the transmission line. Below we consider only the quadrature signals.

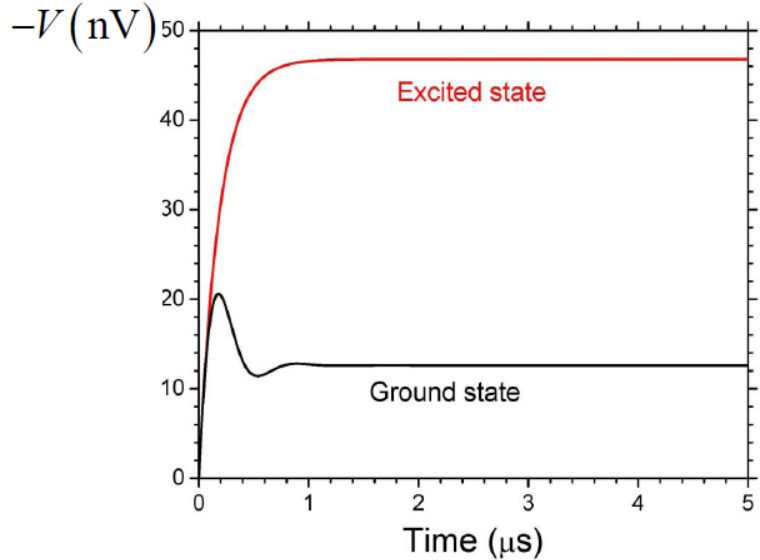

Fig. 2. The negative of the quadrature voltage $(-V)$ at the output of MR for $T_1 = \infty$.

In Figs 2 and 3 we show the negative of the non-stationary quadrature voltage $(-V)$ for parameters that are close to the experimental parameters in [17]:

$$f_r = 6.19 GHz, \ \kappa/2\pi = 1.7 MHz, \ \chi/2\pi = 0.7 MHz.$$

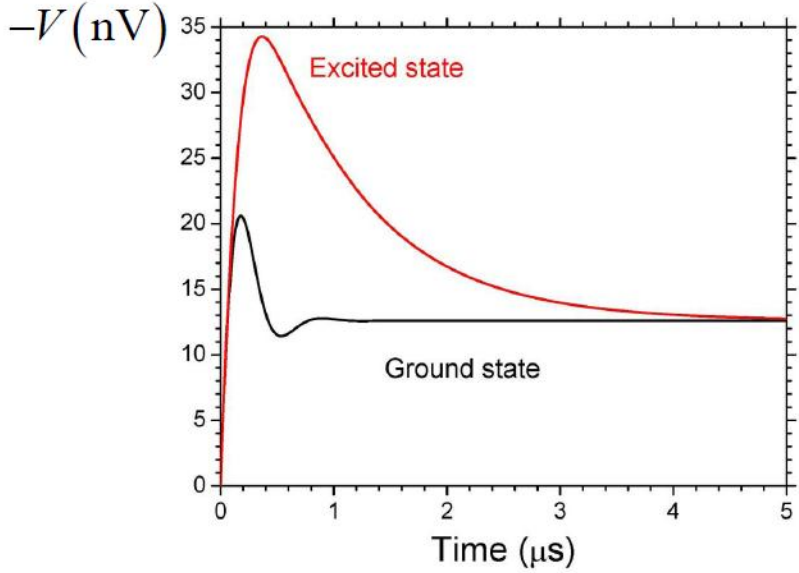

Fig. 3. The same as in Fig. 2 for $T_1 = 900$ns.

Next, we consider amplification of the quadrature signal with the MSA. In the MSA an input coil has one end open, so the parasitic capacitances become active components of the input circuit. As a result, the theoretical description of the input circuit is not straightforward, and the suitable equivalent circuit for the MSA is unknown. Various theoretical approaches to describe the MSA have been suggested in the literature. (See, for example, [16, 19-23].) We will use here a semi-empirical model [16], which provides a reliable estimate for the MSA gain. In this model, one considers a relatively simple MSA scheme consisting of the input circuit and the SQUID. (See Fig. 4.) The trick is that the effective parameters of the scheme are to be found from experiment assuming that the back reaction of the SQUID on the input circuit can be ignored.

With a harmonic input voltage, $V_{in}(t) = V \exp(i\omega t)$, the current in the input coil is given by $I_i(t) = I_i \exp(i\omega t)$. The complex amplitude of the current, $I_i$, can be expressed in terms of the impedances: $I_i = V Z_{LCR} / Z_L Z$, where

$$Z = R_1 + Z_{C1} + Z_{LCR}, \qquad (6)$$
$$Z_{LCR} = (Z_L^{-1} + Z_C^{-1} + R^{-1})^{-1}.$$

The time-dependent input flux from the coil to the SQUID is $MI_i(t)$. For a dispersive measurement, the MSA input frequency equals the MR driving frequency: $\omega = \omega_m$.

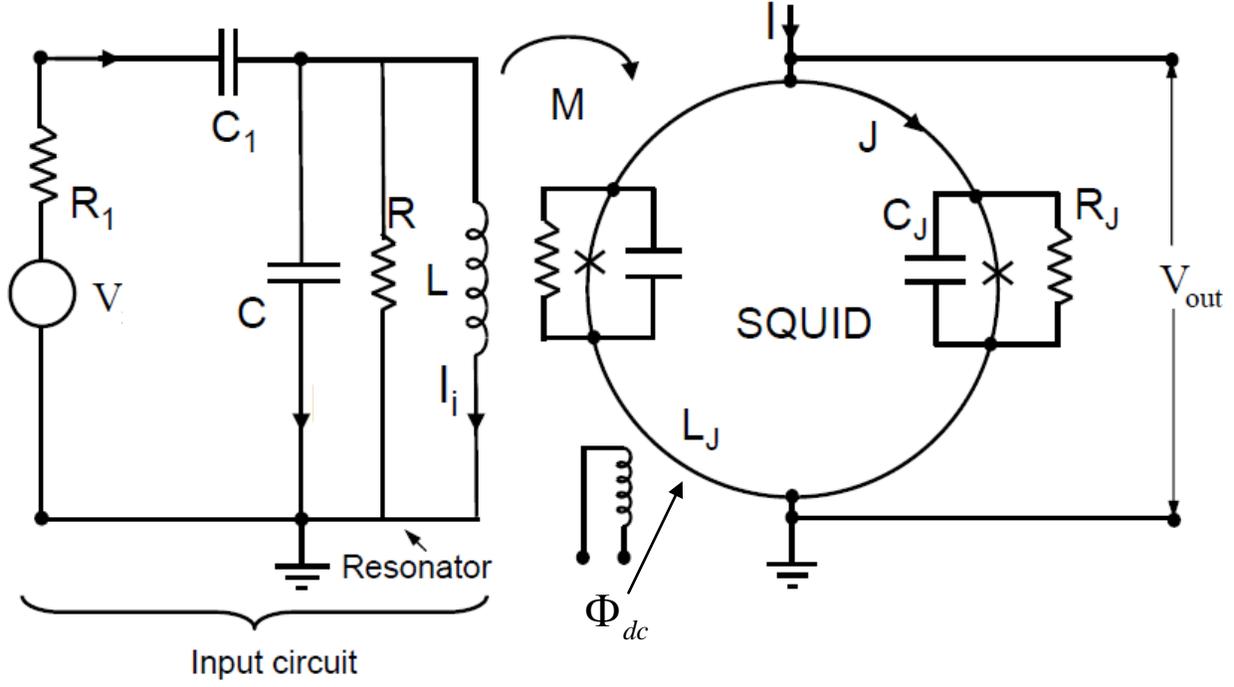

Fig. 4. The MSA equivalent circuit.

In order to compute the output MSA voltage, $V_{out}(t)$, we have solved numerically the standard system of equations for a dc SQUID [24]:

$$\varphi_0 C_J \ddot{\delta}_1 + \frac{\varphi_0}{R_J} \dot{\delta}_1 = \frac{I}{2} - J - I_0 \sin \delta_1$$
$$\varphi_0 C_J \ddot{\delta}_2 + \frac{\varphi_0}{R_J} \dot{\delta}_2 = \frac{I}{2} + J - I_0 \sin \delta_2 \quad (7)$$
$$\varphi_0 (\delta_1 - \delta_2) = \Phi_{dc} + L_J J + MI_i$$
$$V_{out}(t) = \frac{\varphi_0}{2} \left( \dot{\delta}_1 + \dot{\delta}_2 \right)$$

Here the dot above $\dot{\delta}_1$ and $\dot{\delta}_2$ indicates time differentiation; $\varphi_0 = \hbar/2e$ is the reduced flux quantum; $\delta_1$ and $\delta_2$ are the phase differences across JJ; $I_0$ is the JJ critical current; I is the bias current; J is the circulating current; and $\Phi_{dc}$ is the dc bias flux.

Solving Eqs (7), we compute the output MSA voltage, $V_{out}(t)$, and then find its Fourier component at the input frequency, ω. Below we use the symbol, $V_{out}$, for the Fourier component of the output MSA voltage. (The ratio, $V_{out}/MI_i$, represents the dynamical transfer function of the SQUID.) In Fig. 5 we show the power gain $G = |V_{out}/V|^2$ (in dB) as a function of the input frequency, $f = \omega/2\pi$, for the following values of parameters:

$$R_1 = 50\Omega,\ C_1 = 0.12\text{pH},\ L = 0.69\text{nH},$$
$$C = 0.85\text{pF},\ R = 1\text{k}\Omega,\ M = 0.22\text{nH}, \qquad (8)$$
$$I_0 = 8\mu\text{A},\ R_J = 20\Omega,\ L_J = 0.129\text{nH},\ C_J = 52.7\text{fF}.$$

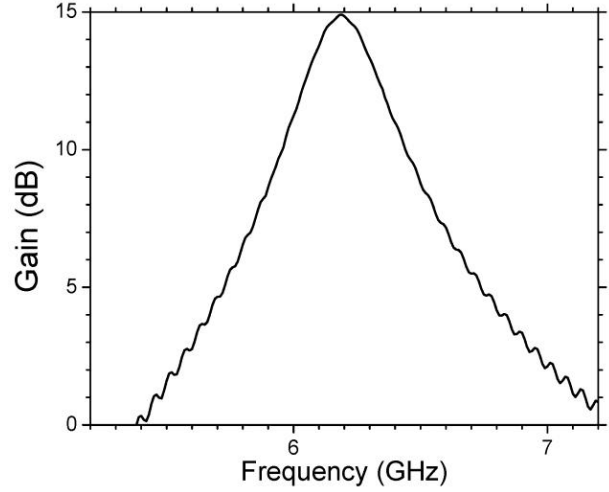

Fig. 5. The MSA power gain, $10\log G$.

For the chosen parameters, the maximum power gain, which we denote as $G_{MSA}$, occurs at the frequency, $f = f_{MSA} = 6.19\text{GHz}$. The value of the maximum gain is 14.9dB. The MSA bandwidth, $B_{MSA} = 340\text{MHz}$ (from 6.02 to 6.36 MHz, where gain decreases by 3dB).

Next, we have computed the power spectral density of the MSA noise at the temperature, $T = 15\text{mK}$, taking into consideration the Johnson noise generated by all the resistors in Fig. 4. In the input circuit in Fig. 4 we set $V_{in} = 0$, and added the two independent sources of voltage noise in series with the resistors, $R_1$ and $R$. In the SQUID circuit in Fig. 4 we have added the two independent sources of current noise in parallel to the resistors, $R_J$. (The voltage noise source in series is equivalent to the current noise source in parallel.) For every realization of the Johnson noise we have solved the system of equations for the input circuit and SQUID and found the corresponding realization of the output noise voltage, $V_{out}^n(t)$. The voltage noise produced by a resistor with resistance, $R_k$, was approximated as white noise with the power spectral density (see, for example, [25]):

$$S_V = 2R_k h f_{MSA} \coth\left(\frac{hf_{MSA}}{2k_B T}\right). \tag{9}$$

The corresponding power spectral density for the current noise is, $S_V / R_k^2$. In order to simulate the white noise we have used a standard continuous chain of the short rectangular pulses with the random amplitude and zero average value. (See, for example, [24].) The power spectral density of the MSA noise at the input frequency, $f$, was computed using the expressions:

$$S_V(f) = \lim_{t_i \to \infty} \frac{2}{t_i}\left\langle \left|V(f)\right|^2\right\rangle, \quad V(f) = \int_0^{t_i} V_{out}^n(t)\exp(2\pi i f t)dt. \tag{10}$$

The spectral density was computed, averaging over 300 realizations of $V_{out}^n(t)$. The maximum of the spectral density, $S_V(f)$, which we denote as $S_{MSA}$, was found at the same frequency, $f = f_{MSA}$, as the maximum of the gain. For our parameters we have obtained the value: $S_{MSA} = 5 \times 10^{-20} \text{V}^2/\text{Hz}$. The MSA noise temperature, $T_n$, was found from equation [26]:

$$2R_1 h f_{MSA} \coth\left(\frac{hf_{MSA}}{2k_B(T_n+T)}\right)G(f_{MSA}) = S_{MSA}. \tag{11}$$

We have obtained the value, $T_n \approx 440\text{mK} \approx 1.5 hf_{MSA}/k_B$. Assuming that after amplification the measured signal is passing through a filter with bandwidth, $B = 2\text{MHz}$, we obtain the voltage noise power, $S_{MSA}B = 10^{-13}\text{V}^2$.

Next, we estimate the SNR for a single-shot dispersive measurement with the MSA. Fig. 6 shows the difference, $D$, between the MSA output voltages, $V_{out}$, corresponding to the two qubit states for $T_1 = 900\text{ns}$. The values of $V_{out}$ were obtained from the voltages shown in Fig. 3 using the computed MSA gain, $G_{MSA}$, from Fig. 5. The function, $D(t)$, has the maximum value, $D_{max} = 120\text{nV}$, at time $t = 0.46\mu\text{s}$. The maximum value of the SNR can be estimated as,

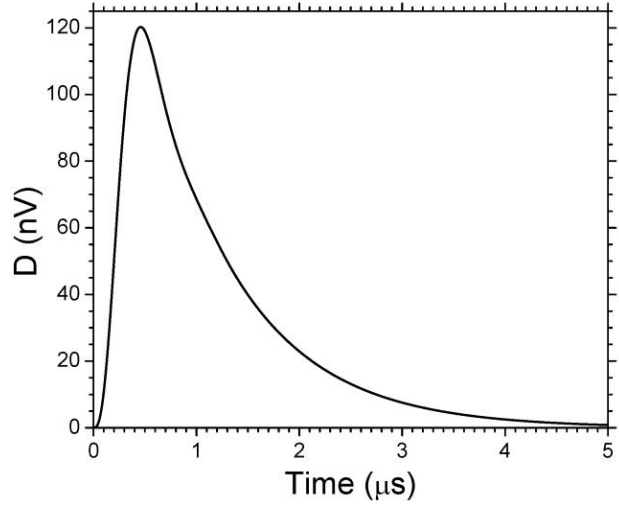

Fig. 6. The difference between the voltages corresponding to the two qubit states.

$$SNR = \frac{D_{max}}{\sqrt{S_{MSA}B}} = 0.425. \qquad (12)$$

The MSA gain can be raised by increasing the resistance, $R_J$, in the MSA SQUID. However, our simulations show that the MSA noise grows faster than the gain, so the SNR drops. In order to obtain $SNR > 1$, we have to increase the input voltage in the MSA by a factor of $K > 2.35$. This means that the average number of photons in the resonator must be $\langle n \rangle = K^2 > 5.5$. Our computations show that the corresponding maximum input voltage in the MSA (about 108 nV) remains in the linear range of the MSA, which extends to about $30\mu\text{V}$.

## 3. Single-shot non-demolition measurement of the phase qubit

In this section we find the expression for the MR-QL coupling constant for the phase qubit. Then, we formulate a condition on the MR-QL mutual inductance that allows a single-shot non-demolition dispersive measurement of the phase qubit with the MSA. We will consider a coplanar waveguide MR inductively coupled to the QL. Fig.7 shows the MR and the QL and their equivalent circuits.

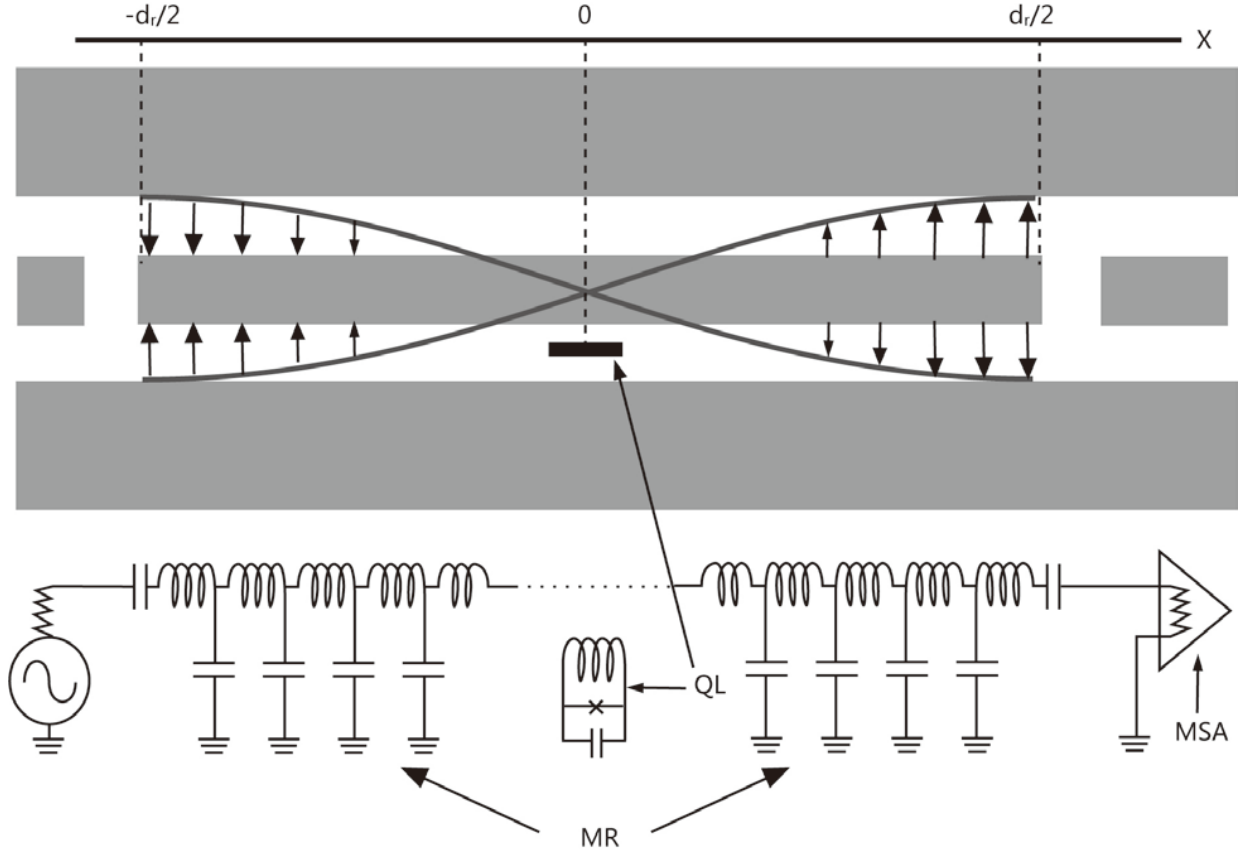

Fig. 7. The MR and the QL and their equivalent circuits. Arrows in the MR show the electric field for the first harmonic; $d_r$ is the MR length; symbol "×" denotes the JJ.

The current, $j(x,t)$, for the first (half wavelength) harmonic in the MR can be described in terms of the creation and annihilation operators [10]:

$$j(x,t) = i(\hbar\omega_r / L_r)^{1/2} \cos(\pi x / d_r)(a^\dagger - a), \qquad (13)$$

where $L_r$ is the MR inductance, and $d_r$ is its length. (See Fig. 7). We assume that the QL is placed near the center of the MR, so we can put $x = 0$. The Hamiltonian of the phase qubit is described by the expression [27]:

$$H_q = p^2/2m + U(\delta, t), \qquad (14)$$

where

$$p = -i\hbar \partial / \partial \delta,$$
$$m = \varphi_0^2 C_q,$$
$$U(\delta) = E_J \left\{ [\delta - \varphi_p - \varphi_1]^2 / 2\lambda_q - \cos \delta \right\},$$
$$\lambda_q = L_q / L_0,$$
$$L_0 = \varphi_0 / I_0,$$
$$\varphi_1 = \Phi_1 / \varphi_0,$$
$$\varphi_p = \Phi_p / \varphi_0.$$
(15)

Here $\delta$ is the phase difference across the JJ, $L_q$ and $C_q$ are the QL inductance and capacitance; $E_J$ is the Josephson energy, $\Phi_p$ is the permanent bias flux on the QL, $\Phi_1 = M_{qr} j$ is the flux produced by the MR on the QL, $M_{qr}$ is the MR-QL mutual inductance, $I_0$ is the JJ critical current.

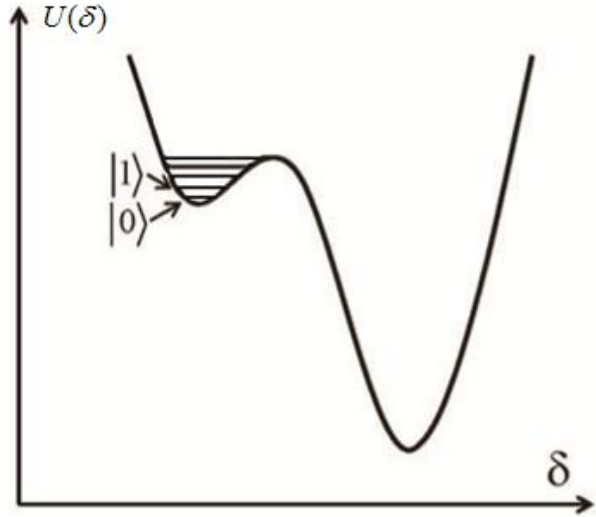

Fig. 8. A double well potential energy of the QL. The phase qubit is spanned over two lowest levels in the shallow well.

The function, $U(\delta)$, in (15) describes a double well potential with a shallow well and a deep well. (See Fig. 8.)

We have taken the values of the QL parameters from experiment [28]:

$$C_q = 700 fF, \ L_q = 720 pH, \ I_0 = 1.7 \mu A. \tag{16}$$

With no MR ($\Phi_1 = 0$) we have adjusted numerically the bias permanent flux, $\Phi_p$, to obtain five levels in the shallow well. (See Fig. 8.) The corresponding qubit frequency is, $f_q = \omega_q / 2\pi = 8.66 GHz$. From Eqs (14) and (15) we obtain the Hamiltonian of the QL-MR interaction:

$$H_{int} = -(E_J / \lambda_q) \varphi_1 \delta. \tag{17}$$

The truncated Hamiltonian of the phase qubit can be written in the form:

$$\begin{pmatrix} H_{ee} & H_{eg} \\ H_{ge} & H_{gg} \end{pmatrix}, \tag{18}$$

where, $|e\rangle$ and $|g\rangle$, are the qubit excited and ground states. In matrix notation:

$$|e\rangle = \begin{pmatrix} 1 \\ 0 \end{pmatrix}, \ |g\rangle = \begin{pmatrix} 0 \\ 1 \end{pmatrix}. \tag{19}$$

The truncated Hamiltonian of the QL-MR interaction can be written in terms of the Pauli operators:

$$H_{int} = \frac{H_{eg} + H_{ge}}{2} \sigma_x + i \frac{H_{eg} - H_{ge}}{2} \sigma_y. \tag{20}$$

Assuming that $H_{eg}$ is real, and taking into consideration Eqs. (17), (15), and (13), we can write the truncated interaction Hamiltonian in the form:

$$H_{int} = -i \frac{M_{qr} E_J}{\lambda_q \varphi_0} \sqrt{\frac{\hbar \omega_r}{L_r}} \langle e | \delta | g \rangle (a^\dagger - a) \sigma_x. \tag{21}$$

Ignoring rapidly oscillating terms, we obtain:

$$H_{int} = -i \frac{M_{qr} E_J}{\lambda_q \varphi_0} \sqrt{\frac{\hbar \omega_r}{L_r}} \langle e | \delta | g \rangle (\sigma^- a^\dagger - \sigma^+ a) = i\hbar g(\sigma^- a^\dagger - \sigma^+ a),$$

$$\hbar g = -\frac{M_{qr} I_0}{\lambda_q} \sqrt{\frac{\hbar \omega_r}{L_r}} \langle e | \delta | g \rangle. \tag{22}$$

In this approximation, the total qubit-resonator Hamiltonian is the Jaynes-Cummings Hamiltonian (see, for example,[10]):

$$H = \hbar \omega_r \left( a^\dagger a + 1/2 \right) + (\hbar \omega_q / 2) \sigma^z + i\hbar g(\sigma^- a^\dagger - \sigma^+ a). \tag{23}$$

The dispersive approximation of the Jaynes-Cummings Hamiltonian is:

$$H = \hbar(\omega_r + \chi\sigma^z)a^\dagger a + (\hbar/2)(\omega_q + \chi)\sigma^z,$$
$$\chi = g^2/\Delta_{qr}, \qquad (24)$$
$$\Delta_{qr} = \omega_q - \omega_r.$$

This approximation is valid if, $4g^2(\langle n \rangle + 1) \ll \Delta_{qr}^2$. The ratio, $(\Delta_{qr}/2g)^2 = \langle n_c \rangle$, defines the critical number of photons in the dispersive approximation [10].

The dispersive Hamiltonian (24) describes the dispersive frequency shift in the MR. For our parameters, $\chi/2\pi = 0.7\text{MHz}$, $f_q = 8.66\text{GHz}$, and $f_r \approx f_{MSA} = 6.19\text{GHz}$, the qubit-resonator coupling should be $g/2\pi = 41.6\text{MHz}$. The critical number of photons for these parameters is, approximately: $\langle n_c \rangle = 880$. Using Eq. (22) and approximating the phase qubit states with the eigenfunctions of the harmonic oscillator, we obtain: $g = (M_{qr}/L_q)(f_r/2f_q L_r C_q)^{1/2}$. For a given value of $g$, the required value of the mutual QL-MR inductance can be found from the expression:

$$M_{qr} = gL_q \sqrt{\frac{2f_q C_q L_r}{f_r}}. \qquad (25)$$

If the MR inductance, $L_r$, takes values between $1nH$ and $10nH$, then the required value of the mutual inductance, $M_{qr}$, ranges from $8.3pH$ to $26\text{pH}$.

**Conclusion**

In this work we studied the possibility of a single-shot non-demolition measurement of a superconducting qubit using an MSA. We have shown that, for reasonable values of parameters, this measurement is possible with about six photons in the MR. For a phase qubit inductively coupled to the MR we estimated the required value of the QL-MR mutual inductance. Our results can be useful for experimental implementation of a single-shot non-demolition measurement of the superconducting qubit.

**Acknowledgement**

the U.S. Department of Energy at Los Alamos National Laboratory under Contract No. DEAC52-06NA25396 and by Lawrence Livermore National Laboratory under Contract DE-AC52-07NA27344, and was funded by the Office of the Director of National Intelligence (ODNI), and Intelligence Advanced Research Projects Activity (IARPA). All statements of fact, opinion or conclusions contained herein are those of the authors and should not be construed as representing the official views or policies of IARPA, the ODNI, or the U.S. Government.

**References**

1. M.H. Devoret and J.M. Martinis. Implementing qubits with superconducting integrating circuits. Quantum Information Processing, **3**, 163 (2004).
2. J.Q. You and F. Nori. Superconducting circuits and quantum information. Phys. Today, **58**, 42 (2005).
3. J. Majer, J.M. Chow, J.M. Gambetta, J. Koch, B.R.Johnson, J.A. Schreier, L. Frunzio, D.I. Schuster, A.A. Houck, A. Wallraff, A. Blais, M.H. Devoret, S.M. Girvin, and R.J. Schoelkopf. Coupling superconducting qubits via a cavity bus. Nature, **449**, 443 (2007).
4. J. Clarke and F.K. Wilhelm. Superconducting quantum bits. Nature, **453**, 1031 (2008).
5. L. DiCarlo, J.M. Chow, J.M. Gambetta, L.S. Bishop, B.R. Johnson, D.I. Schuster, J. Majer, A. Blais, L. Frunzio, S.M. Girvin, and R.J. Schoelkopf. Demonstration of two-qubit algorithms with a superconducting quantum processor. Nature, **460**, 240 (2009).
6. M. Neeley, R.C. Bialczak, M. Lenander, E. Lucero, M. Mariantoni, A.D. O'Connell, D. Sank, H. Wang, M. Weides, J. Wenner, Y. Yin, T. Yamamoto, A.N. Cleland, J.M. Martinis, Generation of three-qubit entangled states using superconducting phase qubits, Nature **467,** 570 (2010).
7. J. Gambetta, W.A. Braff, A. Wallraff, S.M. Girvin, R.J. Schoelkopf. Protocols for optimal readout of qubits using a continuous quantum nondemolition measurement, Phys. Rev. A **76**, 012325 (2007).
8. G.P. Berman, A.R. Bishop, A.A. Chumak, D. Kinion, and V.I. Tsifrinovich. Measurement of the Josephson Junction Phase Qubits by a Microstrip Resonator. arXiv:quant-ph/09123791 (2009).
9. G.P. Berman, A.A. Chumak, D.I. Kamenev, D. Kinion, and V.I. Tsifrinovich. Non-demolition adiabatic measurement of the phase qubit state. Quantum Information & Computation, **11**, 1045 (2011).
10. A. Blais, R.S. Huang, A. Wallraff, S.M. Girvin, and R.J. Schoelkopf. Cavity quantum electrodynamicsfor superconducting electrical circuits: An architecture for quantum computation. Phys. Rev. A **69**, 062320 (2004).
11. M. Mück and R. McDermott, Radio-frequency amplifiers based on dc SQUIDs. Superconductor Science and Technology, **23**, 093001 (2010).
12. M. Mück, J.B. Kucia, and J. Clarke, Superconducting quantum interference device as a near-quantum- limited amplifier at 0.5 GHz. Appl. Phys. Lett. **78**, 967 (2001).
13. M. Boissonneault, J.M. Gambetta, and A. Blais. Nonlinear dispersive regime of cavity QED: The dressed dephasing model. Phys. Rev. A **77**, 060305(R) (2008).

14. G.P. Berman and A.A. Chumak. Influence of external fields and environment on the dynamics of a phase-qubit–resonator system. Phys. Rev. A **83**, 042322 (2011).

15. G.P. Berman, A.A. Chumak, and V.I. Tsifrinovich. Transitional dynamics of a phase qubit-resonator system: requirements for fast readout of a phase qubit. arXiv:quant-ph/1110.5566v1 (2011).

16. D. Kinion and J. Clarke. Microstrip superconducting quantum interference device radio-frequency amplifier: Scattering parameters and input coupling. Appl. Phys. Lett. **92**, 172503 (2008).

17. R. Bianchetti, S. Filipp M. Baur, J. M. Fink, M. Göppl, P. J. Leek, L. Steffen, A. Blais, and A. Wallraff. Dynamics of dispersive single-qubit readout in circuit quantum electrodynamics. Phys. Rev. A **80**, 043840 (2009).

18. J.M. Martinis. Superconducting Phase Qubits. Quantum information processing, **8**, 81 (2009).

19. M. Mück, J. Clarke. The superconducting quantum interference device microstrip amplifier: Computer models. J. Appl. Phys. **88**, 6910 (2000).

20. M. Mück, J.B. Kycia, J. Clarke, Appl. Phys. Lett. **78**, 967 (2001).

21. R. Bradley, J. Clarke, D. Kinion, L. J. Rosenberg, K. van Bibber, S. Matsuki, M. Mück, and P. Sikivie. Microwave Cavity Searches for Dark-Matter Axions. Rev. Mod. Phy*s*. **75**, 777-817 (2003).

22. G.P. Berman, A.A. Chumak, D.I. Kamenev, D. Kinion, and V.I. Tsifrinovich. Modeling and simulation of a microstrip-SQUID amplifier. J. of Low Temp. Phys., **165**, 55, (2011).

23. G.P. Berman, A.A. Chumak, and V.I. Tsifrinovich. The renormalization effect in the microstrip SQUID amplifier. arXiv: quant-ph/ 1110.5579v1 (2011).

24. C.D. Tesche and J. Clarke. dc SQUID: Noise and optimization. J. of Low Temp. Phys., **29**, 301 (1977).

25. D. Abbott, B. R. Davis, N.J. Phillips, and K. Eshraghian. Simple Derivation of the Thermal Noise Formula Using Window-Limited Fourier Transforms and Other Conundrums. IEEE Transactions on Education, **39**, 1 (1996).

26. J. Clarke, A. T. Lee, M. Mück and P. L. Richards. SQUID Voltmeters and Amplifiers. The SQUID Handbook Vol. II Applications of SQUIDs and SQUID Systems, (eds. J. Clarke, A. I. Braginski) Wiley-VCH Verlag GmbH & Co. KGaA, Weinheim, p. 1-93 (2006).

27. A.G. Kofman, Q. Zhang, J.M. Martinis, and A.N. Korotkov. Theoretical analysis of measurement crosstalk for coupled Josephson phase qubits. Phys. Rev. B **75**, 014524 (2007).

28. R. McDermott, R.W. Simmonds, M. Steffen, K.B. Cooper, K. Cicak, K. Osborn, S. Oh, D.P. Pappas, and J.M. Martinis. Simultaneous state measurement of coupled Josephson phase qubits. Science **307**, 1299 (2005).